\documentclass[preprint,showpacs,amsmath,amssymb,pre,superscriptaddress,floatfix]{revtex4-1}

\usepackage{graphicx}
\usepackage{dcolumn}
\usepackage{bm}
\usepackage{amssymb}
\usepackage{multirow}

\newcommand{\1}{\begin{equation}}
\newcommand{\2}{\end{equation}}
\newcommand{\ea}{\begin{eqnarray}} 
\newcommand{\ee}{\end{eqnarray}}
\newcommand{\4}[2]{{\frac{#1}{#2}}}

\newcommand{\de}{{{\rm d}}}

\begin{document}

\title{Neutral particle focusing in composite driven dissipative billiards}

\date{\today}

\author{Benno Liebchen}
\email[]{Benno.Liebchen@physnet.uni-hamburg.de}
\author{Christoph Petri}
\author{Mario Krizanac}
\author{Peter Schmelcher}
\email[]{Peter.Schmelcher@physnet.uni-hamburg.de}
\affiliation{Zentrum f\"ur Optische Quantentechnologien, Universit\"at Hamburg, Luruper Chaussee 149, 22761 Hamburg, Germany}%

\begin{abstract}
Dynamical focusing of ensembles of neutral particles in energy and configuration space has been demonstrated recently [C. Petri et al. 2010, Phys. Rev. E (R) {\bf 82}, 035204] using time-dependent elliptical billiards.
The interplay of nonlinearity, dissipation and driving yields the occurrence of attractors in the phase space of the billiard.
Here we show that dissipative oval billiards with slowly oscillating elliptical scatterers in the interior allow for a dynamical focusing
on simple periodic trajectories with close to perfect efficiency. 
This setup should be more amenable to corresponding experiments which are briefly discussed. 
\end{abstract}

\maketitle
\paragraph*{Introduction}
Exhibiting typical phenomena occurring in nonlinear systems which are too complex to analyze them directly, 
billiards have become popular model systems in many areas of physics:
the main concepts of quantum chaos could be developed in the context of billiards \cite{stoeckmann99} (and refs. therein), 
they stimulated the foundations of modern semiclassics \cite{stoeckmann99} and could be related to the Riemann hypothesis \cite{bunimovich05}. Even a justification of a probabilistic approach to statistical mechanics
can be based on billiards \cite{zalavsky99,egolf00}. Recently, they could be related to the physics of graphene \cite{miao07} and have been exploited in order to improve the efficiency of thermoelectric materials \cite{casati08}. 
Meanwhile, they are experimentally accessible in a variety of setups such as
microwave resonators \cite{stoeckmann90,graef92,stoeckmann99}, atom-optical- \cite{milner01,friedman01,montangero09}, and semi-conductor heterostructure based devices \cite{weiss91} 
as well as quantum dots \cite{marcus92}.
\\In recent years particularly time-dependent billiards attracted much attention. A related key topic is the long-standing quest of the general requirements of the occurrence of Fermi-acceleration \cite{blandford87,lichtenberg92,karlis06} (and refs. therein), 
i.e. under which conditions does a system 'heat up' unboundedly due to a driving force?
The difficulty of a general answer becomes already clear by the exemplary findings, that correlations can either cause the suppression of Fermi-acceleration, as in  
the one-dimensional Fermi-Ulam model \cite{lichtenberg92,liebermann72,lichtenberg80} or can even yield exponential acceleration \cite{gelfreich08,shah10,gelfreich11,liebchen11}. 
\\Due to their conceptual simplicity and the complexity of their phenomenology, time-dependent billiards are perfectly suited to employ ideas which are based on characteristic phenomena of complex systems:
In \cite{petri10} the fact that the set of attractors in the phase space of a dissipative nonlinear systems determines its possible asymptotic behavior
was employed to an elliptical billiard with a time-dependent boundary in order to achieve a filter for neutral particles, working simultaneously in momentum and configuration space. 
Cutting a small hole into the boundary a synchronized particle beam is obtained. 
Such a dynamical particle filter, based on general concepts of nonlinear dynamics 
provides a simple, versatile and efficient alternative to the extensively used standard filtering-devices such as lenses, nozzles, mirrors or collimators.
However, the corresponding results were achieved using a dissipative elliptical billiard with oscillating hard boundaries which are experimentally challenging to realize
and turn out to lead to relatively high loss rates in terms of particles stopping inside the billiard. 
\\Static boundaries are comparatively simple to realize and design and lead in combination with a time-dependent scatterer in the interior of
the billiard to an alternative dynamical system. This raises the question, whether particle focusing as demonstrated in \cite{petri10} can be obtained also in such a setup.
A further limitation of \cite{petri10} concerns the relatively high particle losses due to rotators in the phase space of elliptical billiards \cite{petri10}. 
\\The main goal of this work is to provide and analyze a model which combines experimental simpleness with high-efficiency asymptotic focusing and synchronization of random ensembles of neutral particles on a single periodic trajectory.
A billiard with oval boundaries and oscillating elliptical scatterer in the interior is shown to provide the appropriate setup and demonstrated to yield the desired focusing effects for sufficiently slow driving and an elongated shape of the inset. 
\paragraph*{Motivation of the Model}
It is first discussed why for the class of billiards with a static boundary and a time-dependent scatterer in the interior, 
an oval for the former and an ellipse for the latter appears to be the simplest promising setup in order
to achieve a highly efficient particle focusing. Then, a corresponding model is proposed. 
The obvious starting point would be to replace the breathing elliptical boundaries in \cite{petri10} by static ones and to place a circular, oscillating scatterer in 
the interior of the billiard. 
However, due to the presence of rotators in the elliptical billiard \cite{lenz10} relatively high particle losses would be the consequence 
which can be only avoided by introducing local defects at the boundary \cite{petri10}. 
Thus, the natural extension is to consider oval boundaries which in the following turn out to be perfectly tunable to avoid particle losses. 
Concerning the geometry of the inset, the natural starting point of a circular inset turns out to result either in attractors of chaotic type, i.e. in a non-synchronized, non-periodic asymptotic dynamics
or in very low focusing efficiencies. These considerations therefore result in the following model: 
\begin{figure}[htb]
\begin{center}
\includegraphics[width=0.9\textwidth]{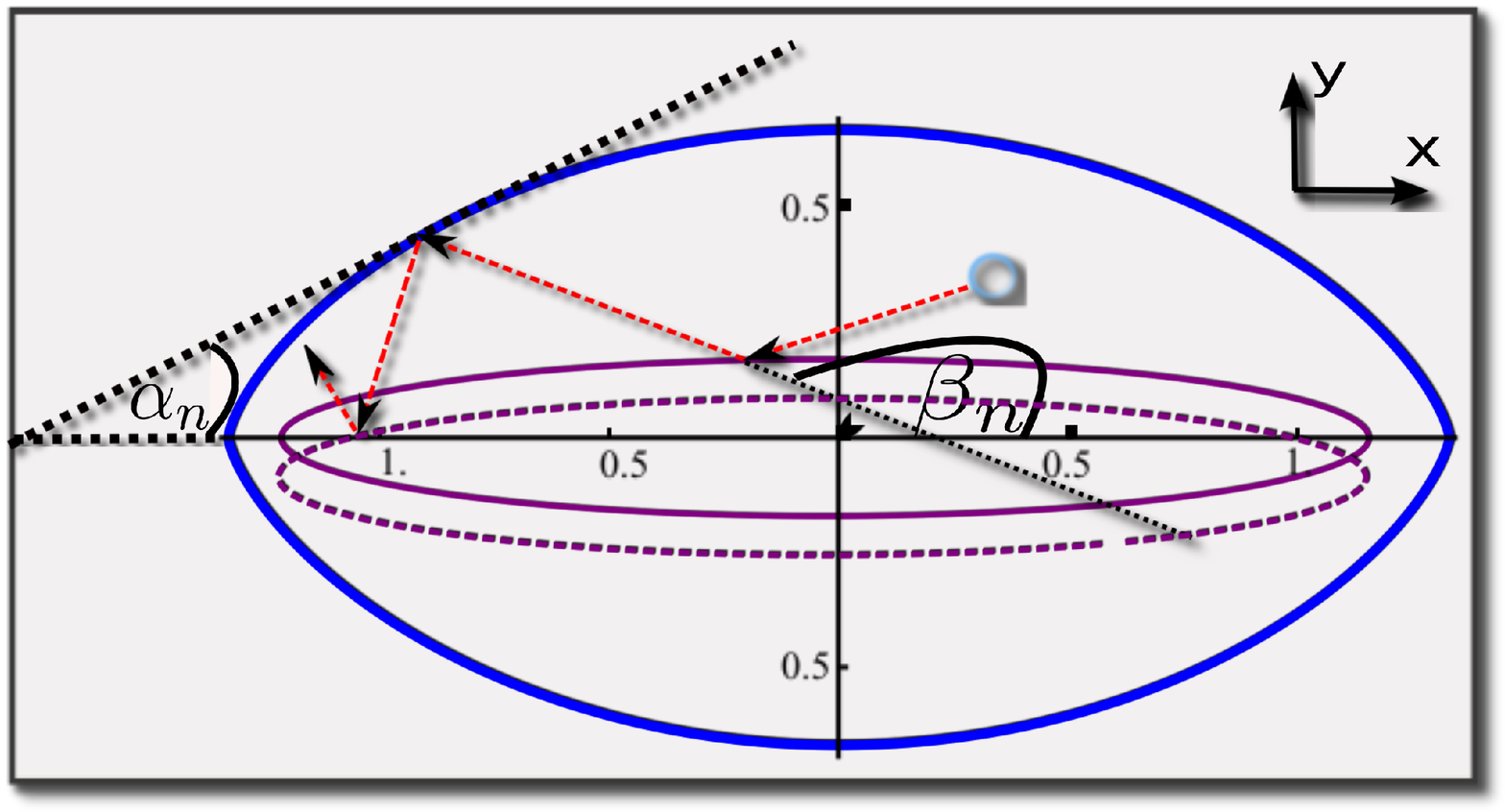} 
\caption{Sketch of the oval billiard with the oscillating inset in its interior at two different collision-times (full and short dashed purple line), a typical trajectory (red, long dashed)
and the angles $\alpha_n,\beta_n$ for an exemplary collision with the static boundary.}
\label{fig0}
\end{center}
\end{figure}
Consider the dynamics of non-interacting point particles of mass $m$
in a two-dimensional static oval billiard with an oscillating hard scatterer of elliptical shape in the interior (Fig.~\ref{fig0})
in the presence of a velocity-proportional friction force $F_R=k \cdot {\bf v}$. 
The static oval boundary is defined as a plain, closed curve ${\bf x}_B(\phi)$ parametrized by $0 \leq \phi < 2\pi$ 
\cite{berry81}: 
\ea
x_B(\phi) &=& \left[(\delta/2 +1) \sin\phi + (\delta/6)\sin(3\phi) \right] \nonumber  \\ 
y_B(\phi) &=& \left[(\delta/2-1)\cos\phi-(\delta/6)\cos(3\phi) \right]  
\label{boundaries}
\ee
Therein $0 \leq \delta <1$ is a measure for the deviation of the oval from the circle (which is reobtained for $\delta = 0$) and for the maximal curvature, 
occurring at $\phi=0,\pi$.
The respective boundary of the oscillating inset can be defined directly by the following coordinate representation
\ea
&& \4{[x_I-f_x(t)]^2}{A^2} + \4{[y_I-f_x(t)]^2}{B^2}=1 \nonumber \\
&& f_x(t)=a\cos(\omega_x t);\;\; f_y(t)=b\cos(\omega_y t) \label{inset},
\ee
where $A,B$ are the semi-major and semi-minor axes of the ellipse and 
$\omega_x,\omega_y,a,b$ denote the frequencies and amplitudes of the oscillation in $x,y$-direction, respectively.  
\\The corresponding free-space dynamics of a particle with coordinates ${\bf x}\equiv(x,y)$, initial position ${\bf x}_0$ and velocity ${\bf v}_0$ 
is described by the Newtonian equation of motion $m {\ddot {\bf x}} + k {\dot {\bf x}}=0 $,
with the solution 
\1 \label{freespacesolution} 
{\bf x}(t)={\bf x}_0 +  \4{m {\bf v}_0}{k} \left(1-  {\rm e}^{-(k/m) t}\right)
\2
Thus, particles inside the billiard move on straight lines between successive collisions with boundary/inset and their dynamics
can be captured by a mapping from one collision to the next one. 
Collisions can of course happen in any order with the boundary and inset and not only in an alternating way. 
Particles which do not hit the driven inset for a sufficiently long time eventually stop and are considered as lost. 
\\In order to minimize losses, the occurrence of rotors in the phase space, corresponding to particles traveling around the inset without hitting it, must be suppressed. 
Accordingly, in the following a large deviation of the oval from the circle ($\delta=0.99$) and mainly elongated insets are chosen (cf.~\ref{fig0}). 
\paragraph*{Mapping:}
A mapping which describes the dynamics of particles in the billiard from collision to collision is formulated in the following. 
Let $n$ count the total number of collisions of the particle with the boundary/inset and $t_n,{\bf x}_n,{\bf v}_n$ 
denote time, position and velocity 
at the instance after the corresponding collision. 
Then, the dynamics of the system can be described by a five-dimensional, non-area-preserving and non-Hamiltonian 
mapping 
\1 \label{mapping} \mathcal{M}:(t_n,{\bf x}_n,{\bf v}_n)\mapsto (t_{n+1},{\bf x}_{n+1},{\bf v}_{n+1}) \2
Since it is not clear a priori whether the respective next collision happens with the boundary or with the inset, both are calculated and the smaller one is selected.
In the following the iteration of the mapping is discussed for the possible cases, collisions with the oval boundary and the elliptical inset.
\\\emph{Collision with the oval boundary:}
First on, the slope of the trajectory after the respective last collision is calculated, then position and velocity and finally the corresponding time of the next collision follow. 
Let $\alpha_n$ measure the angle between the tangent to the boundary at the $n$-th collision point 
and the $x-$axis and $\beta_n$
the angle between the velocity vector and the $x$-axis at this collision point (cf. \ref{fig0}). Then, 
the slope of the tangent at the $n$-th collision point $m_{B,n}$ can be calculated by 
\1 m_{B,n} = \left.\4{\de y_B(\phi)}{\de x_B(\phi)}\right|_{\phi=\phi_n} = \left.\4{\de y_B/\de \phi}{\de x_B/\de \phi}\right|_{\phi=\phi_n}
= \4{(1-\delta/2)\sin\phi + (\delta/2)\sin(3\phi)}{(1+\delta/2)\cos\phi + (\delta/2)\cos(3\phi)} \2
and $\alpha_{n} = \arctan(m_{B,n})$ holds. 
With $\beta_{n+1}=2\alpha_n - \beta_n$ the trajectory 
after the $n$-th collision has the slope $m_{T,n}:=\tan(\beta_{n+1})$, i.e. it evolves along a straight line with a known slope of $y(x)= m_{T,n} (x-x_n) + y_n $.
Inserting the boundary-functions $x_B(\phi)$ and $y_B(\phi)$ (Eq.~\ref{boundaries}) for $x$ and $y$ yields 
an implicit equation determining the value of the angle 
at which the particle hits the boundary (in the inset-free) case at the respective next collision:
\ea \label{boundarycolls} g(\phi) :&=& m_{T,n} \left[(\delta/2+1)\sin \phi + (\delta/6)\sin(3\phi) \right]\nonumber \\ &-& \left[(\delta/2-1)\cos\phi - (\delta/6)\cos(3\phi)  \right] 
+ (y_n-m_{T,n} x_n) = 0 \ee
This implicit equation determines the boundary parameter $\phi_{n+1} \equiv \phi$ corresponding to the respective next collision point of particle and oval and
can be solved numerically with a standard Newtonian root finding method. 
Inserting $\phi_{n+1}$ into Eq.~(\ref{boundaries}) yields the collision points in configuration space
$x^B_{n+1}=x_B(\phi_{n+1})$ and $y^B_{n+1}=x_B(\phi_{n+1})$.
Therein the superscript $B$ denotes the collision with the static boundary.
With Eq.~\ref{freespacesolution} and ${\bf x}_{n+1}-{\bf x}_n$ being parallel to $ {\bf v}_n$, the corresponding collision time results as 
$t^B_{n+1} = (-m/k) \ln\left[1- k|{\bf x}_{n+1}-{\bf x}_n|/(m v_n) \right] +t_n$. The new velocity 
${\bf v}^B_{n+1}$ directly follows with $v^B_{n+1}=v_n \cdot {\rm exp}\left[(-k/m) (t_{n+1}-t_n)\right]$ and $v^y_{n+1}/v^x_{n+1}=\tan(\beta^{n+1})=m_{T,n}$.
\\\emph{Collision with the oscillating elliptical inset:}
In contrast to the oval boundary, for calculating the collision time with the inset, advantage can be taken from the fact, that a 
coordinate representation for the ellipse exists. 
Inserting Eq.~(\ref{freespacesolution}) with ${\bf v}_0 \mapsto (v_{x,n},v_{y,n})$ and $t \mapsto t_{n+1}-t_n$ component-wise into Eq.~\ref{inset}
yields the following implicit equation determining the time $t_{n+1}$ of the respective following collision with the inset: 
\ea \label{insetcolls} 
&&\left[x_n + \frac{v_{x,n} m}{k} (1-\rm{e}^{- \frac{k}{m}(t_{n+1}-t_n)}) -f_x(t_{n+1})\right]^2/A^2 \nonumber \\ 
&+& \left[y_n+ \frac{v_{y,n} m}{k} (1-{\rm e}^{-\frac{k}{m}(t_{n+1}-t_n)}) - f_y(t_{n+1}) \right]^2/B^2 -1 \nonumber \\
 &=&:f(t_{n+1}) = 0; \quad t_{n+1}-t_n > 0 \ee
Note, that Eq.~(\ref{insetcolls}) only has a finite solution when $f(t_{n+1})$ changes sign at least once in the interval
$(t_n,t_{n+1}^B)$. In this case, the respective smallest solution determines $t_{n+1}$ and the respective next collision occurs with the inset.
Otherwise $t_{n+1}=t^B_{n+1},{\bf x}_{n+1}={\bf x}^B_{n+1}, {\bf v}_{n+1}={\bf v}^B_{n+1}$ is chosen. 
For a collision with the inset, the velocity changes as 
\1 \label{veloupd}
{\bf v}_{n+1} = {\bf v}_n {\rm e}^{\left[(-k/m) (t_{n+1}-t_n)\right]} -2 \left[{\bf n}\cdot \left({\bf v}_n{\rm e}^{\left[(-k/m) (t_{n+1}-t_n)\right]} - \delta {\bf v}_I(t)\right)\right] \cdot {\bf n}
\2
where 
$\delta {\bf v}_I(t)=\left.\partial_t {\bf x}_I (t)\right|_{t=t_{n+1}}$ is the velocity vector of the inset at the respective collision point and ${\bf n}=\delta {\bf v}_I(t)/|\delta {\bf v}_I(t)|$ is the corresponding normal vector.
After each collision the slope can be directly updated via $m_{T,n+1}=v^y_{n+1}/v^x_{n+1}$ such that the respective part can be skipped above after a collision with the inset. All considerations concerning the time of 
the respective next collision of course hold independently of the sequence (boundary/inset) in which the collisions occur and the mapping can be straightforwardly iterated. 
\paragraph*{Particle Focusing}
The mapping is numerically iterated for ensembles of $10^4$ initial conditions for $2.8\cdot 10^6$ collisions and various choices of inset geometry and driving law.
For the following discussion four representative examples are chosen. 
Initial velocities are chosen randomly with uniform probability in $v \in [5,20]$ (the lower offset is to avoid particles stopping before hitting the inset 
for the first time). 
\begin{figure}[htb]
\begin{center}
\includegraphics[width=0.7\textwidth]{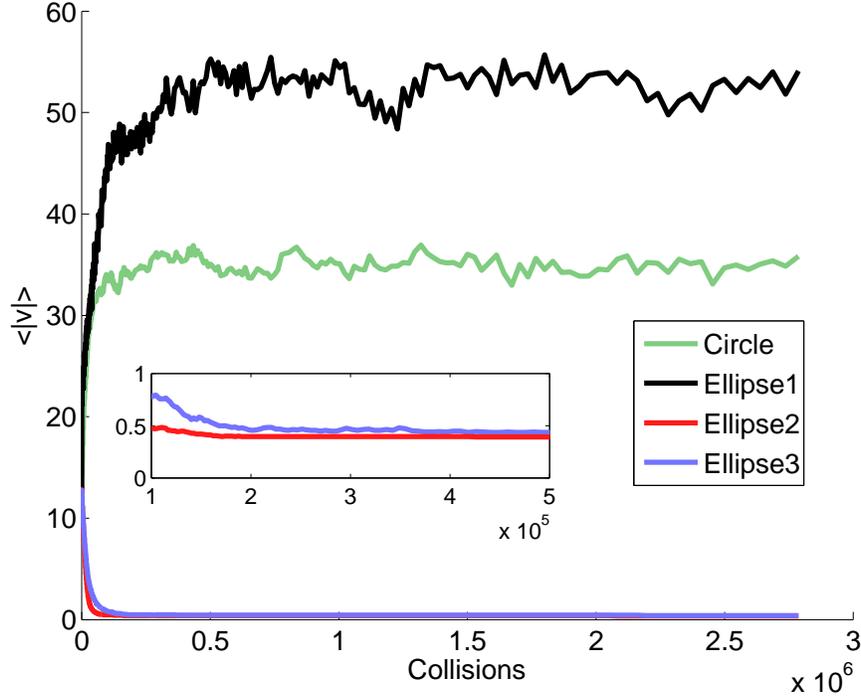} 
\caption{Time-evolution of the ensemble-averaged particle velocity.
Parameters $\omega_x=\omega_y=1.0$: Circle: $A=B=0.5;\;a=b=0.1;\; k=0.002$; Ellipse 1: $A=1.0; B=0.25; a=0;\; b=0.25;\; k=0.002$; Ellipse 2: $a=1.3; b=0.05; A=0; B=0.05;\; k=0.001$;
Ellipse 3: $a=1.2; b=0.05; A=0.1; B=0.05;\; k=0.001$.}
\label{fig1}
\end{center}
\end{figure}
Regarding the evolution of the ensemble averaged magnitude of the velocity $<|v_n|>$, according to Fig. \ref{fig1}, saturation is 
observed for all four visualized combinations of geometry and driving law. 
This saturation suggests that all particles have approached attractors, i.e. that the system is in its asymptotic state at the end of the simulation-time.
For the circular inset (circle) and the relatively fast driven elliptical inset (ellipse 1) the average-velocity grows initially, then
saturates at a certain value well above the average initial velocity. 
In contrast to that the average velocity decreases for both slowly driven elliptical billiards (ellipse 2,3) 
and converges towards values of $<|v_n|> \sim 0.4$.
Comparatively strong driving (for example $A=B=0.3$ for the circular inset - not shown) leads to unbounded growth of energy
on the examined timescales and is thus inappropriate in order to achieve particle focusing. 
\begin{figure}[htb]
\begin{center}
\includegraphics[width=0.7\textwidth]{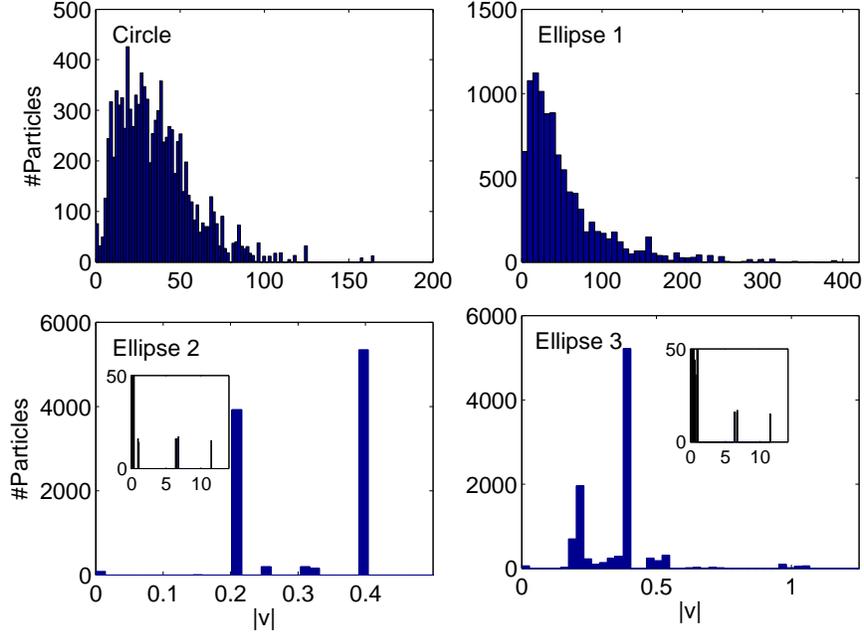} 
\caption{Velocity-distributions after $2.8 \cdot 10^6$ collisions. Parameters like in Fig.~\ref{fig1}.}
\label{fig2}
\end{center}
\end{figure}
\begin{figure}[htb]
\begin{center}
\includegraphics[width=0.6\textwidth]{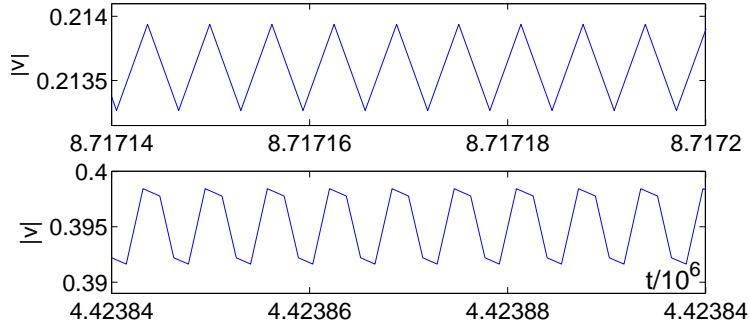} 
\caption{Time-dependent magnitude of the velocity after $2.8 \cdot 10^6$ collisions for the configuration of ellipse 2 for two different initial conditions. Parameters like in Fig.~\ref{fig1}.}
\label{fig3}
\end{center}
\end{figure}
\\Now the asymptotic dynamics is analyzed in more detail. 
Fig.~\ref{fig2} (Circle) shows that the slowly driven circular inset
leads to a broad asymptotic velocity distribution.
Thus, it is inappropriate in order to achieve a focusing of particles on single trajectories. 
The broad distribution together with the observed saturation of the mean ensemble velocity is indicative of the presence of a strange attractor, i.e. that 
trajectories are asymptotically focused onto a submanifold with lower (and non integer) dimension than the phase space.
An obvious conclusion would be to assume that circular insets do not allow for the presence of isolated attractive 
periodic orbits since defocussing is stronger than the focusing of the oval boundary in regions 
where attractors are conceivable. Thus, elliptical insets of elongated shape should be more promising. 
However, the velocity distribution (Fig.~\ref{fig2}, ellipse 1) shows that 
even the combination of a very flat elliptical inset and moderate driving is not sufficient in order to achieve particle focusing: 
again a stationary, i.e. non-spreading,
but broad velocity distribution and corresponding chaotic motion of the respective trajectories is obtained.
\\A slowly and purely vertically driven flat elliptical inset provides an asymptotic velocity profile with two 
clearly pronounced peaks at $|v| \sim 0.2$ and $|v| \sim 0.4$
containing $92.7\%$ of all initial conditions (Fig.~\ref{fig2}, ellipse 2). 
Less than one percent of all particles are lost (low peak at $|v| = 0$).
The peak at $|v| \sim 0.2$ corresponds to a periodic orbit attractor of period two. The corresponding asymptotic time-evolution of the magnitude of the velocity is shown in  
Fig.~\ref{fig3} (upper plot).  
The amplitude of the oscillations of $|v(t)|$ continuously decreases as the attractor is approached.
The related motion in configuration space is a two-periodic bouncing 
exactly on the $y$-axis between inset and boundary. 
About $39.3\%$ of all initial conditions are asymptotically focused on this periodic orbit and are thereby synchronized to perform asymptotically one and the same dynamics.
The peak at $|v| \sim 0.4$ corresponds to a periodic orbit attractor of period four (Fig.~\ref{fig3}, lower plot) which splits into four individual peaks in a very fine 
resolved version of Fig.~\ref{fig2} (ellipse 2) which merge in the long-time limit. 
About $53.4\%$ of all initial conditions are focused on this periodic trajectory and are thereby perfectly synchronized in the long-time limit. 
In configuration space, the corresponding particles are focused on the same orbit as for the $|v| \sim 0.2$ peak.
The remaining particles are focused on a variety of more complicated periodic trajectories, all in the velocity regime
$0.2<|v|<0.39$ and very few ones on a trajectory around $|v| \sim 5$ (Inset in Fig.\ref{fig2}, ellipse 2.).
This demonstrates, that it is indeed possible to achieve the focusing of randomly distributed ensembles of neutral particles on periodic 
trajectories with close to perfect efficiency by using static oval billiards with elongated, slowly driven elliptical insets.  
\\It is now suggestive that the respective, purely vertically aligned periodic orbit can also become attractive 
for a circular inset but with a much smaller basin of attraction. 
Since the circle is stronger defocussing than the boundary is focusing 
sophisticated time-correlations are required in order to approach it. 
Corresponding numerical simulations confirmed this expectation:
for sufficiently slow driving, circular insets also yield attractive periodic trajectories and may thus be used in principle for particle focusing, 
but the corresponding efficiencies are 
extremely poor. The best achieved efficiency was well below $10\%$, i.e. typically most particles stop before reaching periodic orbits.
\\Further numerical simulations have been performed for a variety of combinations of slow driving laws and elongated elliptical insets and always yielded
both, the occurrence of fixed point attractors in phase space and small loss rates. 
Consequently the demonstrated particle focusing does not require any fine tuning of parameters. 
Fig.~\ref{fig2} (Ellipse 3) exemplarily shows the resulting velocity distribution for a slowly driven, vertically oscillating inset.
Again, more than half of the ensemble is focused asymptotically onto the above-discussed period-four attractor, but the rest of 
the ensemble asymptotically follows a variety of periodic orbits with velocities between $|v|\sim 0.18$ and $|v|\sim 0.55$.
A very different velocity profile could be achieved for example for $a=1.0; b=0.05; A=0.3; B=0.0125; k=0.001; \omega_x=2.5; \omega_y=0.5$
but only for the price that a relevant part of the ensemble is asymptotically governed by a chaotic attractor and thus remains non-synchronized. 
\\Note that the particle-focusing mechanism is governed solely by the topological structure of the phase space and thus works independently of the 
specific choice of the initial conditions. Key ingredients for high efficiency focusing are strong curvature of the oval boundary at certain positions
and elongated insets such that particles are trapped in the upper or in the lower half of the billiard for many collisions.
\\Finally, possible experimental implementations are discussed. The presented numerical results should be accessible to particle-clusters in a micromechanical 
realization of the billiard-system with medium. Alternatively, a purely optical verification of the mechanism should be equally possible by using atoms in 
laser-based optical traps of oval shape, counter-propagating laser fields yielding friction due to the optical molasses and an oscillating blue detuned laser taking the
role of the oscillating, reflecting inset. According to the following consideration the parameter sets used in this work should be effectively accessible in both setups:
In Eqs.(\ref{freespacesolution}),(\ref{boundarycolls}),(\ref{insetcolls}),(\ref{veloupd}) the length scale occurs only in the driving amplitudes $a,b$ and in $a \omega_x, b\omega_y$, the time-unit appears in the 
driving frequencies $\omega_x,\omega_y$, in $a \omega_x, b\omega_y$ and in the friction coefficient, while the mass only appears reciprocal to the friction coefficient which scales with mass and time scale.
A desired friction coefficient can be effectively realized by choosing particles with appropriate mass or 
also via the frequency of the driving, but the latter determines the size of the billiard, which otherwise can be chosen arbitrarily, as long as 
$a \omega_x, b\omega_y$ are preserved. 
\paragraph*{Conclusion}
It has been demonstrated, that dynamical focusing of ensembles of neutral particles in energy and configuration space, is possible in billiards with static boundaries and an oscillating scatterer in the interior. 
Employing appropriate oval boundaries and sufficiently slowly driven elongated elliptical insets, close to perfect efficiency can be achieved. 
The suggested focusing device can be implemented either in a micromechanical or in a purely optical setup yielding a versatile alternative to common focusing devices like lenses or collimators.
\paragraph*{Acknowledgements}
B.L. thanks the Landesexzellenzinitiative Hamburg "`Frontiers in Quantum Photon Science"'
which is funded by the Joachim Herz Stiftung for financial support.

\end{document}